# Circular Currents in Molecular Wires

Dhurba Rai, Oded Hod and Abraham Nitzan
School of Chemistry, Tel Aviv University, Tel Aviv 69978, Israel.

## Abstract

We consider circular currents in molecular wires with loop substructures studied within simple tight-binding models. Previous studies of this issue have focused on specific molecular structures. Here we address several general issues. First we consider the quantitative definition of a circular current and adopt a definition that identified the circular component of a loop current as the sole source of the magnetic field induced in the loop. The latter may be associated with the field at the loop center, with the magnetic moment associated with this field or with the total magnetic flux threading the loop. We show that all three definitions yield an identical measure of the loop current. Secondly, we study dephasing effects on the loop current and the associated induced magnetic field. Finally, we consider circular currents in several molecular structures: benzene, azulene, naphthalene and anthracene and show that circular currents occur generically in such structures and can be, in certain voltage ranges, considerably larger than the net current through the molecule, and are furthermore quite persistent to normal thermal dephasing.

# 1. Introduction

While most studies of transport properties of molecular conduction junctions have focused on the overall conduction properties associated with given junction geometry and electronic structure,[1-2] some attention has been given to current distribution within the junction. Such studies have come to the forefront recently with several papers addressing possible interference effects resulting from the existence of multiple conduction pathways.[3-11] In another context, we have recently studied current transfer processes, in which a current imposed on one pathways affects a current in another,[12] and their manifestation in affecting efficiencies and yields of charge transport in helical molecular structures.[13] Because interference plays a central role in such processes, understanding the role played by relaxation and dephasing is an important related issue.

An interesting phenomenon encountered when addressing the current distribution within the molecular framework connecting the junction metallic leads is the possibility to induce circular currents.[14-21] Observations of such phenomena are so far limited to theoretical computations on model molecular junctions, but calculations done on several different systems yield broadly consistent results: First, circular currents often appear in certain voltage regimes in junctions characterized by multiple pathways that may close within a given molecular bridge to give a circular pathway. Second, in some voltage regimes the circular currents can be considerably larger than the net junction current.[15-16,19] Third, such strong circular currents appear near conduction thresholds in the current-voltage characteristic that are associated with nearly degenerate pairs of molecular orbitals whose contribution to the net current is rendered small by destructive interference. In the isolated ring these orbitals are degenerate, and are characterized by equal and opposite orbital angular momentum along the molecular ring.[16,19,21-22] Finally, such circular currents are found to be associated with considerable magnetic fields at the center of the ring.[17,20,23] While several suggestions were made for possible experimental demonstration of the existence of such currents,[17,23] to the best of our knowledge no such experiments were reported so far.

Circular currents in molecular rings as well as in other ring conductors have been discussed in two other contexts. The possibility to excite such currents using circularly polarized light have been indicated by theoretical calculations.[24-27] Persistent currents in mesoscopic conducting rings, that have been under discussion since their prediction in 1983 by Büttiker, Imry and Landauer,[28] are induced by an external magnetic field. It was discovered, however, that closely related loop currents can be induced in the absence of external magnetic fields in rings driven by an external voltage bias[29-32] or by an external electromagnetic field[33] Indeed, such circular currents are closely related in nature to those discussed above for the molecular ring systems.

In spite of many discussions of circular currents in these different contexts, a unique definition of such currents has not been given. Consider the two terminal junction displayed in Fig. 1. The net total current in the external leads is $I_{tot}$ and the currents in the two arms of the ring are $I_1$ and $I_2$. Many of the papers cited above discuss the circular current only qualitatively, identifying the occurrence of a circular current as the case where the segmental currents $I_1$ and $I_2$ have similar signs, so that the magnitude of the current in at least one segment is larger than $|I_{tot}|$. A quantitative definition has been suggested in Ref. 29 where circular currents have been associated only with such situations, identifying the circular current component as the smaller of $(|I_1|,|I_2|)$. Such a definition seems to us rather arbitrary.

In this paper we reconsider the issue of circular currents with three objectives. First, we suggest an alternative quantitative measure of the circular current in ring coupled to an arbitrary number of external leads. Second, we examine the effect of dephasing processes, always to be expected in molecular junctions which are usually studied at room temperature, on these circular currents. Such processes are expected to be important in molecular junctions that are usually studied at room temperature. Finally, we use this new understanding of circular currents to re-examine, within simple tight binding (Hückel) level calculations of the type considered previously in such studies, the magnitudes of the circular currents and the associated induced magnetic fields that are expected in molecular junction involving simple molecular junctions. In a subsequent paper we will examine the way in which the presence of such circular currents is manifested in the interaction of such molecular ring structures with an external magnetic field.

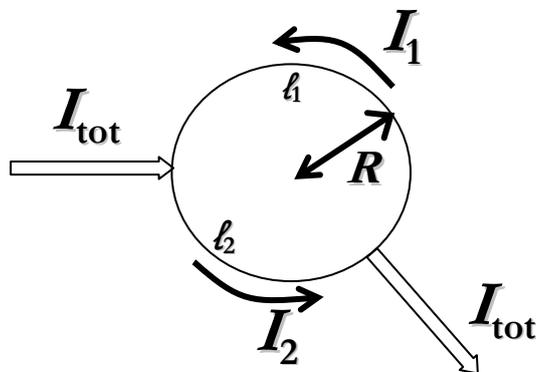

Figure 1. Current distribution in a two terminal junction with a circular ring connecting to conducting leads. The current in any segment of the ring is defined to be positive when it flows in the counter-clockwise direction.

## 2. Circular currents in rings with external links

Consider the system of Figure 1, where a current flows between two leads through a ring of radius $R$. The overall junction current is denoted $I_{tot}$ and the currents in the two ring segments between the leads are $I_1$ and $I_2$. A positive sign is assigned to current flowing in the counter-clockwise direction. Obviously, with this sign convention, any decomposition of the currents in the ring segments into a circular component $I_c$ and transverse components $I_1^{tr}$ and $I_2^{tr}$ satisfies

$$I_{tot} = I_2 - I_1 = (I_2 - I_c) - (I_1 - I_c) = I_2^{tr} - I_1^{tr} \tag{1}$$

We propose to make the choice of $I_c$ unique by assigning it to be the sole source of current induced magnetic field threading the ring. Putting differently, the transverse components, $I_1^{tr}$ in ring segment 1 and $I_2^{tr}$ in ring segment 2 are defined such that their combined contribution to this magnetic field vanishes.

It is not obvious that even this requirement defines the circular and linear components of the ring current uniquely. Indeed we could request that the total magnetic field at the ring center due to $I_1^{tr}$ and $I_2^{tr}$ vanishes, or that the corresponding magnetic moment vanishes, or, most generally, that the total magnetic flux threading the ring due to $I_1^{tr}$ and $I_2^{tr}$ is zero. In appendix A we show that, in fact, all these measures lead to an identical definition of the circular and transverse components of the ring current, as follows:

$$I_c = \frac{1}{L}(I_1 l_1 + I_2 l_2) \tag{2}$$

$$I_1^{tr} = -I_{tot} \frac{l_2}{L}; \quad I_2^{tr} = I_{tot} \frac{l_1}{L} \tag{3}$$

where $l_1$ and $l_2$ are the arc lengths of the corresponding ring segments and where $L = l_1 + l_2 = 2\pi R$ is the circumference of the ring. Note that $I_1^{tr}$ and $I_2^{tr}$ flow in the same direction of $I_{tot}$; the appearance of a negative sign in expression (3) for $I_1^{tr}$ results from the sign convention defined above. It is interesting to note that if the ring is homogeneous, so that it's classical Ohm's law resistance $R_j$ satisfies for any ring segment $j$ $R_j = \alpha l_j$ we have in this classical limit $I_1 l_1 = -I_2 l_2$ which implies that $I_c = 0$. The existence of a circular current under these circumstances is thus seen to be a purely quantum phenomenon.

The considerations that lead to Eq. (2) can be generalized in two ways (see Appendix A). When the ring is replaced by a regular (i.e., cyclic and equilateral) polygon of $n$ sides, Eqs. (2) and (3) remain valid, and may be also represented by $I_c = n^{-1}(I_1 n_1 + I_2 n_2)$ and $I_1^{tr} = -I_{tot} n_2/n$; $I_2^{tr} = I_{tot} n_1/n$ where $n_j$ is the number of sides associated with segment $j$ and $n_1 + n_2 = n$. More significantly, if the ring is linked to external leads in $N$ sites so that it is divided into $N$ segments carrying different currents $I_j$, Eq. (2) becomes:

$$I_c = \frac{1}{L}\sum_{j=1}^{N} I_j l_j \; ; \qquad L = \sum_{j=1}^{N} l_j \tag{4}$$

Eqs. (2)-(4) are used below to evaluate the bias driven circular currents associated with several molecular ring structures. Before that we outline in the next section the tight binding model used for these estimates and the technique used to compute the total current and the associated circular currents that develop in several molecular junction structures with and without dephasing processes.

## 3. Model and method

We consider a molecule described by a tight-binding Hamiltonian model (site energies $\alpha_M$ and nearest-neighbor coupling $\beta_M$) connecting between two leads represented by infinite 1-dimensional tight-binding chains (site energies and nearest-neighbor coupling $\alpha_K$ and $\beta_K$, $K = L,R$, respectively) that represent metal electrodes (see Fig. 2). The Hamiltonian in the site representation is

$$\hat{H} = \hat{H}_L + \hat{H}_R + \hat{H}_M + \hat{V}_{LM} + \hat{V}_{RM} \tag{5}$$

where

$$\hat{H}_K = \alpha_K \sum_{n \in K} |n\rangle\langle n| + \beta_K \sum_{n \in K} \left(|n\rangle\langle n+1| + |n+1\rangle\langle n|\right) \; ; \quad K = L,R,M \tag{6}$$

$$\hat{V}_{KM} = \beta_{KM}\left(|n\rangle\langle m| + |m\rangle\langle n|\right) \quad ; \quad n \in K, m \in M \quad ; \quad K = L, R \quad (7)$$

and where $\{|n\rangle\}$ is a set of orbitals, assumed orthogonal for simplicity, centered about the atomic sites *n* and assumed to span the Hilbert space required for the description of current conduction through the molecular wire under consideration.

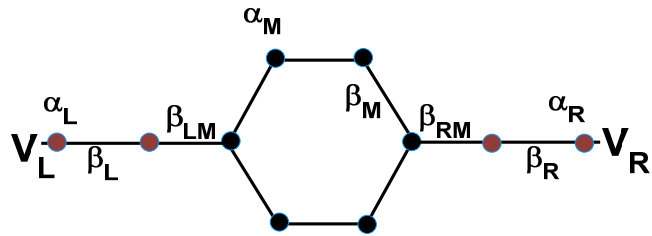

Fig. 2. Tight-binding model for current conduction through a molecule (here represented by Benzene structure connecting between the two 1-dimensional metal leads), $L$ and $R$ with voltage bias $V_L - V_R$.

There are several ways to compute the current distribution within the molecular structure bridging between the conducting leads, and in this paper we adopt the method used in Refs. 12,34. In the amplitude version of this approach we consider a networks of connected sites described by a tight binding Hamiltonian, with a source wire in which electrons are injected into the system and one or more drain wires on which carrier absorption is affected by the exactly known self energy terms. The latter arise from treating explicitly a finite ("interior") system and representing the effect of infinite wires on this system by renormalization of edge sites energies, $E_j \to E_j + \Sigma_j(E)$. $\Sigma_j(E)$ vanishes unless *j* is an edge site on one of the wire segments *K*. In the latter case it takes the form

$$\Sigma_{j \in K}(E) = \frac{(E - \alpha_K) - \sqrt{(E - \alpha_K)^2 - 4\beta_K^2}}{2} \equiv \Lambda_K(E) - (1/2)i\Gamma_K(E) \quad (8)$$

The steady state calculation yields the energy dependent transmission probability $T(E)$ form the source to any drain while at the same time giving the steady state amplitude $C_j(E)$ on each site *j* of the network. The current between any two adjacent sites on the wire segment *K* is then given by

$$I_{K(j-1 \to j)} = \frac{2\beta_K}{\hbar} \text{Im}\left(C_{j-1} C_j^*\right) \quad (9)$$

The density matrix (DM) version of this approach considers a system driven by given DM elements in the incoming wire segment and by absorption terms associated with the current on the outgoing segments, again represented by renormalization of edge site energies. For example, if sites 1 and 2 are located on the incoming wire to the left of the scattering region, the density matrix describing a Bloch wave with wavevector **k** propagating towards and reflected from the scattering region is given in terms of the amplitudes A (taken real) and B of the incident and reflected waves, respectively, by

$$\begin{aligned}
\rho_{11} &= |A|^2; \rho_{22} = |B|^2; \\
\rho_{12} &= |A|^2 e^{-ika} + AB^* e^{ika} \\
\rho_{21} &= |A|^2 e^{ika} + A^* B e^{-ika}
\end{aligned} \quad (10)$$

In the outgoing wire segments, the renormalization of edge site energies by the self energy terms $\Sigma(E)$ appear in the steady state equations for DM elements in the form

$$\dot{\rho}_{jl} = 0 = .... - \frac{1}{2}\left(\Sigma_j(E) + \Sigma_l(E)\right)\rho_{jl} \qquad (11)$$

where "…" represents terms arising from the Hamiltonian (5) written for the interior system and where, again, $\Sigma_j(E)$ vanishes if $j$ is not an edge site.

Pure dephasing in the scattering region (i.e. on the molecular structure) can be included approximately by supplementing the DM equation of motions by phenomenological damping terms associated with phase relaxation. This leads to

$$\dot{\rho}_{jl} = 0 = .... - \frac{1}{2}\left(\gamma_j + \gamma_l\right)\rho_{jl} \qquad (12)$$

where, again, "…" represent all contributions arising from the Hamiltonian (5) and where $\gamma_j$ vanishes unless site $j$ belongs to the molecular bridge. Below we take $\gamma_j = \gamma$, independent of the site on the molecular bridge. The resulting state equations give the amplitude $B$ ($A$ can be taken real with $|A|^2 = f(E)$, where $f(E)$ is the Fermi function associated with the source electrode) as well as the density matrix elements $\rho_{jl}$ associated with all system sites. From these, the outgoing current in any exit wire $K$ is obtained from

$$I_{K(j-1 \to j)} = \frac{2\beta_K}{\hbar}\text{Im}\left(\rho_{j-1,j}\right) = \frac{\Gamma_K(E)}{\hbar}\rho_{jj} \qquad (13)$$

where $j$ is an edge site on wire segment $K$. The equality $I_{j \to l} = (2\beta_{jl}/\hbar)\text{Im}(\rho_{j,l})$ in fact gives the current between any two adjacent sites $j$ and $l$ with intersite coupling $\beta_{jl}$. This yields the overall current, as well as the current through every molecular bond at energy $E$ associated with carriers injected from any given electrode. The current as function of voltage can be obtained by integrating over the energy and summing over all electrodes (the contribution from each electrode is weighted by the corresponding Fermi function. More details of this calculation are provided in Ref. 34.

Finally, the local magnetic field at point $\vec{r}$ inside the molecule is calculated from the Biot-Savart's Law

$$\vec{B}(\vec{r}) = \sum_{(m,n)} \int \frac{\mu_0}{4\pi} I_{m,n} \frac{d\vec{r}' \times (\vec{r} - \vec{r}')}{|(\vec{r} - \vec{r}')|^3} \qquad (14)$$

Where $\mu_0 = 4\pi \times 10^{-7} NA^{-2}$ is the magnetic constant and $\vec{r}'$ is the position vector of an infinitesimal bond current element $I_{m,n} d\vec{r}'$. The summation is taken over all the bonds ($n,m$) inside the molecule.

## 4. Results and discussion

Here, we present computational results obtained for circular currents in a few typical molecules under 'standard' biased junction conditions. The molecular structures chosen have single (benzene) and multiple rings, where in the latter group one may distinguish between separated (biphenyl) and fused (azulene, naphthalene, anthracene) ring structures. It should be emphasized that our calculations, aimed to demonstrate qualitative generic behaviors, use the simplest tight binding models for these structures. Similarly, the electrodes are represented by simple 1-dimensional tight binding chains, each connecting to one specified site of the molecular structure. The results of these calculations should not by any means be considered quantitatively representative, only as indications of typical behaviors. In all calculations we set the on-site energies in the left and right leads to zero, $\alpha_L = \alpha_R = 0$, while the corresponding on-site energies in the molecular structures are taken to be $\alpha_M = $ -1.5 eV. The nearest neighbor coupling parameters are taken to be $\beta_M = 2.5\,\text{eV}$, $\beta_L = \beta_R = 2.4\,\beta_M$ (the latter, unphysically large value is just a way to impose a wide band limit in which we disregard any effect of the finite electrode bandwidth) and $\beta_{LM} = \beta_{RM} = 0.4\,\beta_M$. The leads conduction band are assumed to be half filled, i.e., their zero-bias Fermi energy $E_F$ is taken zero. The imposed potential bias is assumed to fall on the metal-molecule bond, and to be distributed symmetrically between the two molecule-electrode contacts. Thus, the biased electrochemical potentials are $\mu_L = eV/2$ and $\mu_R = -eV/2$. The temperature is taken zero throughout our calculations.

Figure 3 shows the current-voltage characteristics of such model benzene junctions taken as the bridging molecule, for the para, meta and ortho bridging configurations. The effect of dephasing, imposed on the benzene sites as described in Section 3, is shown as well. The effect of multiple pathways on molecular conduction was already discussed in the past[3-11] and we will not dwell on it here. We briefly note that both the geometrical and the dephasing effects on conduction reflect the fact that in the model benzene molecule the molecular orbitals manifested in the observed transport are doubly degenerate; their amplitudes combine differently for different connection schemes and different dephasing rates.

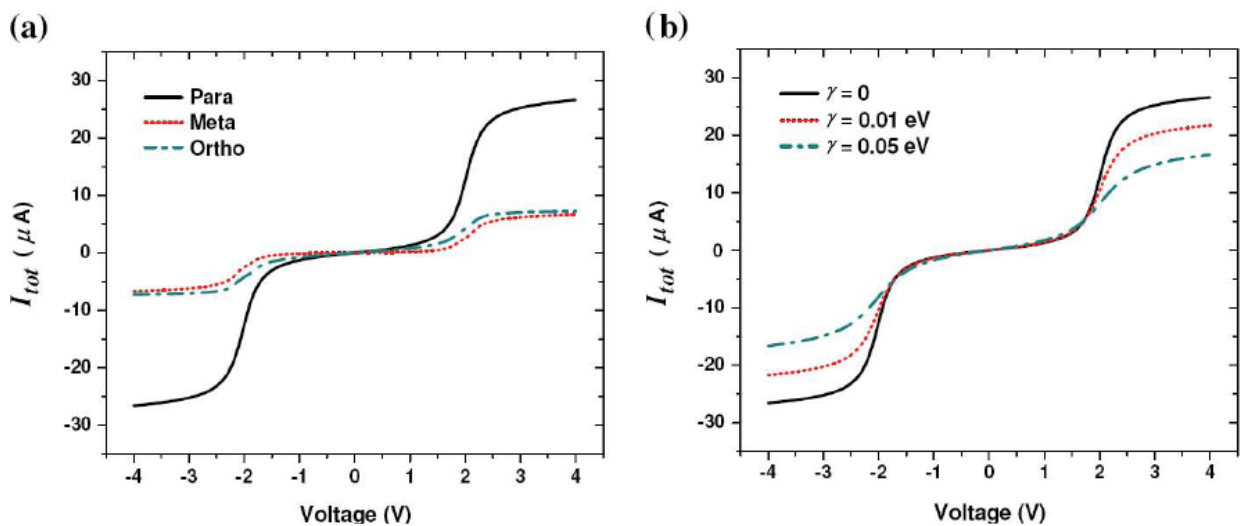

Fig. 3. (a) I-V characteristics of para, meta and ortho-connected benzene. (b) Current-voltage characteristics of para-connected benzene at different dephasing rates ($\gamma$).

The doubly degenerate benzene orbitals can be characterized by their orbital angular momentum, representing Bloch waves going clockwise or counter-clockwise along the ring. Because degeneracy is removed by the molecule-electrode coupling, circular currents arise when one of these waves is expressed more strongly then the other in the conduction, leading to a circular current, a situation that can arise in some voltage ranges in meta- and ortho-connected benzenes, but not in para-connected molecule. This observation may also be described in terms of interference between the two pathways available to an electron moving between the two contacts. The former point of view makes it understandable that the direction of the circular current can depend on the imposed bias, while the latter one suggests sensitivity to dephasing processes. Figures 4 and 5 demonstrate both phenomena. Fig. 4 shows how the directions of the circular current and the associated magnetic field in meta- and para-connected benzene change in different voltage regimes. Fig. 5 shows the magnitude of the circular current and the associated magnetic field at the molecular center as functions of voltage and dephasing rate for these molecules. The sharp resonance features observed can be understood as reflecting the fact that in a relatively narrow voltage regime only one of the two levels associated with opposite orbital angular momenta is in the Fermi window, while in most voltage regimes both contribute, albeit slightly differently because of their split energies.

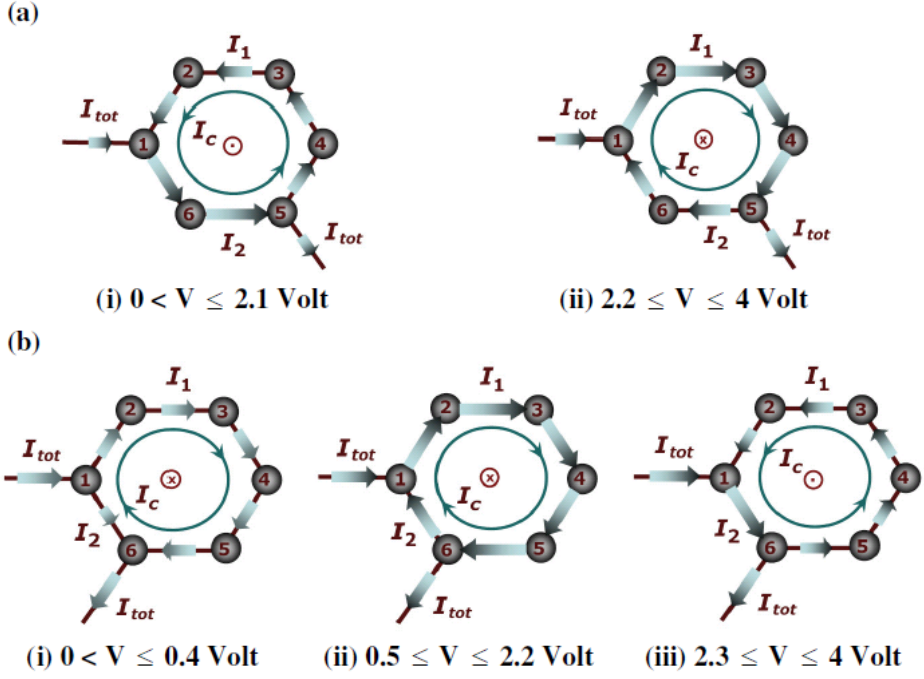

Fig. 4. Internal current distribution in (a) meta- and (b) ortho-connected benzene rings for applied bias in the range 0 to 4V. The blue-green circle depicts circular currents, showing their direction. The direction of the corresponding magnetic fields at the ring centers are shown by encircled dots and crosses representing upward (out of page) and downward (into page) directions, respectively. The arrow sizes indicate the magnitude of the bond currents.

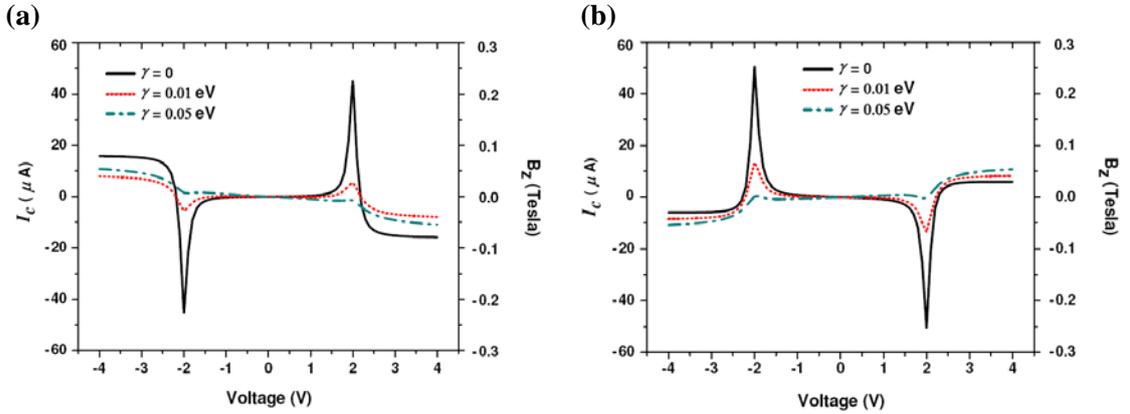

Fig. 5. (a) Variation of circular current ($I_c$, left axis), and magnetic field $B = (0, 0, B_z)$ (right axis) at the center of the meta-connected (panel (a)) and ortho-connected (panel (b)) ring, with applied bias $V$ for different dephasing rates ($\gamma$) A positive circular current corresponds to the counter clockwise direction.

Focusing on the behaviors shown in Figs. 3 and 5, three observations are noteworthy. First, the circular current can be much larger than the total net current carried by the molecule. At resonance, near V ~ 2V, $I_c \approx 18 I_{tot}$ and $I_c \approx 12 I_{tot}$ in the meta- and ortho-connected geometries, respectively. Second, the corresponding induced magnetic field at the ring center is considerable, reaching the maximum of 0.23 Tesla in meta-connected benzene and 0.25 Tesla in the ortho-connected configuration. Finally, both the net current (Fig. 3b) and the circular current (Fig. 5) decrease with increasing dephasing rate on the ring, however the effect of dephasing on the resonance feature of the circular current and the associated magnetic field is considerably stronger than its effect otherwise. Remarkably, the circular current feature is maintained also in the presence of fairly fast dephasing processes.

As was noted in Section 2, our definition of circular current differ from another definition, e.g. Ref 29, where this current component is represented by a reverse (relative to the total) current in one of the ring branches and declares the circular current component to be zero if such reverse current does not exist. A comparison of the two definitions is shown in Fig. 6, where the zero circular current associated with the latter definition is evident in the voltage range $2.5 \leq V \leq 2.8$.

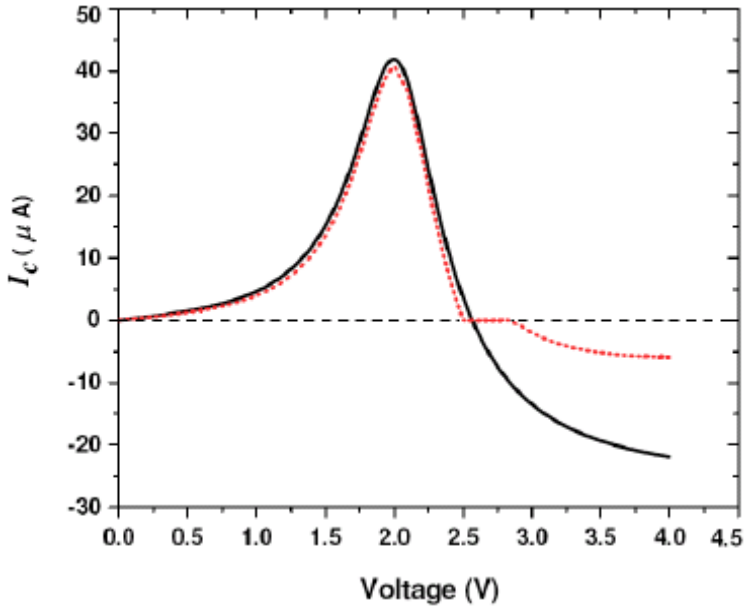

Fig 6. The circular current defined by Eq. (2) (full line; black) and according to Ref. 29 (dotted line; red) plotted as a function of applied bias for the meta-connected structure. Junction parameters are taken as above, except that $\beta_{LM} = \beta_{RM} = 2$ eV.

The strong (relative to the transverse current) circular current that may develop in molecular rings has been noted by several previous authors.[15-16,19] We have noted above that the ratio $|I_c/I_{tot}|$ is affected by the dephasing rate. Interestingly, we find that near resonance the most important parameter affecting this ratio is the molecule-electrode coupling. Fig. 7 shows this trend for the meta-connected benzene structure at bias voltages 2 V (near resonance) and 1 V (off resonance). We note in passing that a circular current is observed also for asymmetric metal-molecule couplings, i.e. $\beta_{LM} \neq \beta_{RM}$ as well. For $\beta_{LM} = 1\,\text{eV}$, although in our calculation the largest circular current was obtained in the symmetric case.

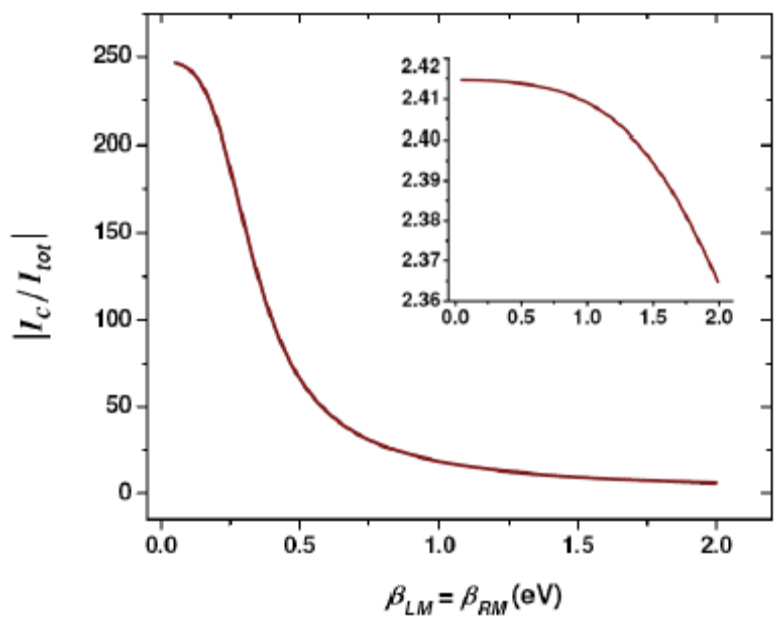

Fig. 7. Variation of current ratio $I_c/I$ with metal-molecule coupling strength at 2 Volt applied bias. The insert shows the same ratio for bias of 1 V.

Observations with other ring structures as bridging molecules are qualitatively similar to those with the benzene structure. Results for the biphenyl structures are shown in Fig. 8. Here, the coupling between the two benzene rings is taken to be same as that between ring sites (2.5 eV). Again, results depend on the connection geometry and no circular current exists in the para (1, 10)-connecting case. A new interesting observation is the fact that in some voltage regime the circular currents on the two rings can be opposite to each other. Fig. 8 shows results obtained for the (2, 11) connection geometry (in a sense, a series of two meta connected benzenes), where the circular currents on the two rings are equal in magnitude and opposite in direction.

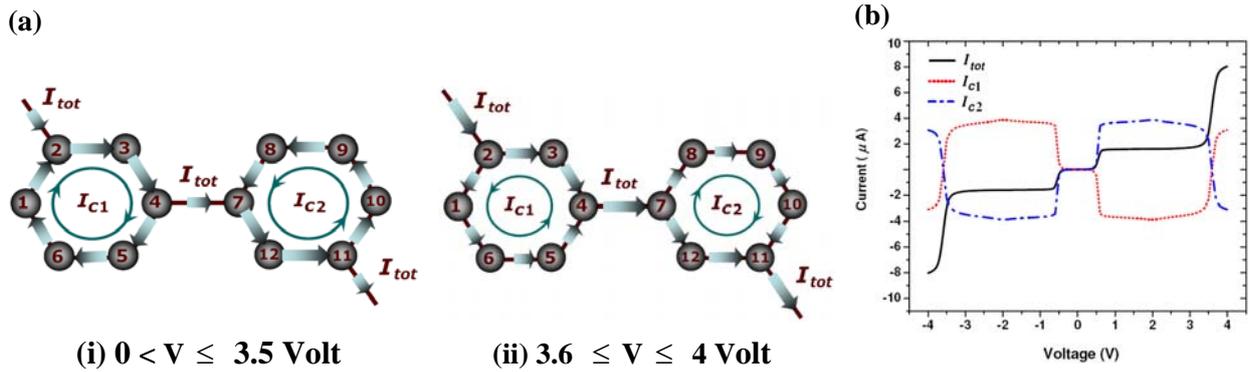

Fig. 8. (a) Internal current distribution pattern in diagonally connected biphenyl. (b) Net current I and circular currents ($I_{c1}, I_{c2}$) as a function of applied bias voltage, V.

Similar results for azulene, naphthalene and anthracene structures are shown in Figures 9-13. Figure 9 shows the behavior of the azulene model. We note that current in both rings are in the same direction, that the circular current in the 5-member ring is larger than that of the 7-member one and that inversion of the circular current direction is not observed in the voltage range 0-4 eV. Obviously, in the symmetrically connected azulene ((1,6) connection, not shown) circular currents do not exist. Symmetry implies that in this case the current on the (4, 8) bond also vanishes for all voltages, a situation reminiscent of balanced Wheatstone's bridge encountered in elementary electrical circuits.

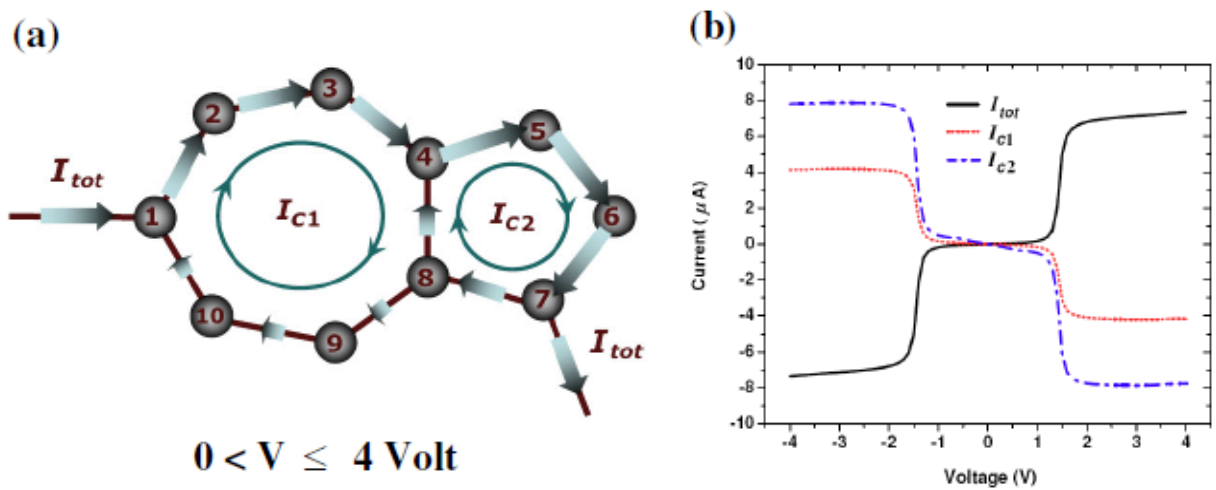

Fig. 9. (a) Circular currents in asymmetrically connected azulene for applied bias in the range 0 to 4 V. (b) Variation of net current I and circular currents ($I_{c1}, I_{c2}$) with the applied bias.

For naphthalene in the (1,6) connection geometry (Fig. 10) the two ring currents are equal and in opposite directions that switch sign at 1.8 V. When compared to the net current, the calculated circular components are relatively small. We attribute this to the fact that the structure is very nearly symmetrically connected. In the (1,7) connection geometry (Fig. 11) the circular currents on the two rings are in the same directions in the voltage range studied, however bond currents can change directions in different voltage regimes as shown. Similar qualitative behaviors are found in the case of anthracene structures (Figs. 12, 13). It is interesting to note that in the structure shown in Fig. 13 no circular current exists in the central benzene ring.

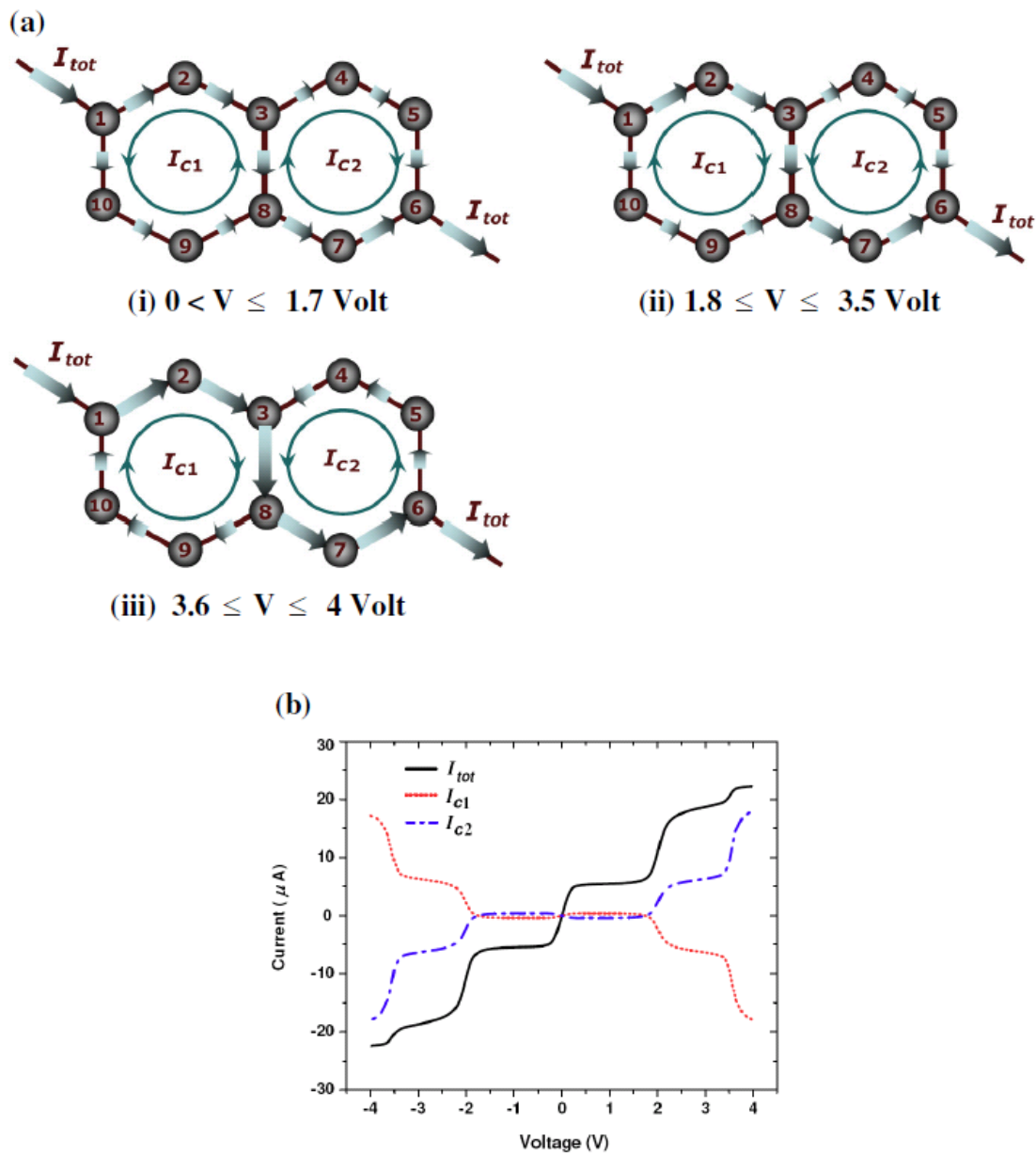

Fig. 10. (a) Current channels in the diagonally connected naphthalene. (b) Variation of net current $I_{tot}$ and circular currents ($I_{c1}, I_{c2}$) with the applied bias.

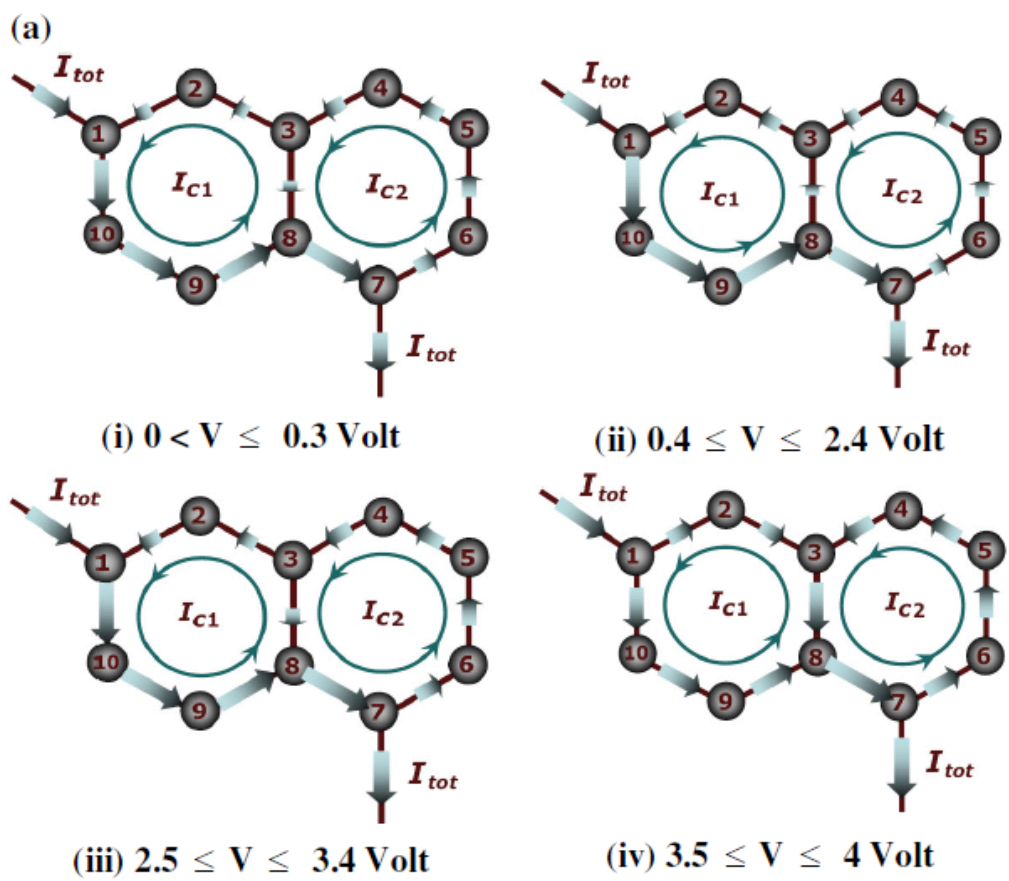

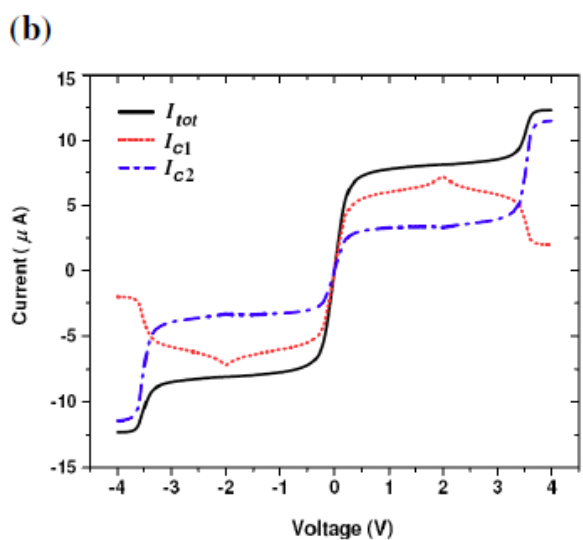

Fig. 11. (a) Circular currents in asymmetrically connected naphthalene. (b) Current-voltage characteristics.

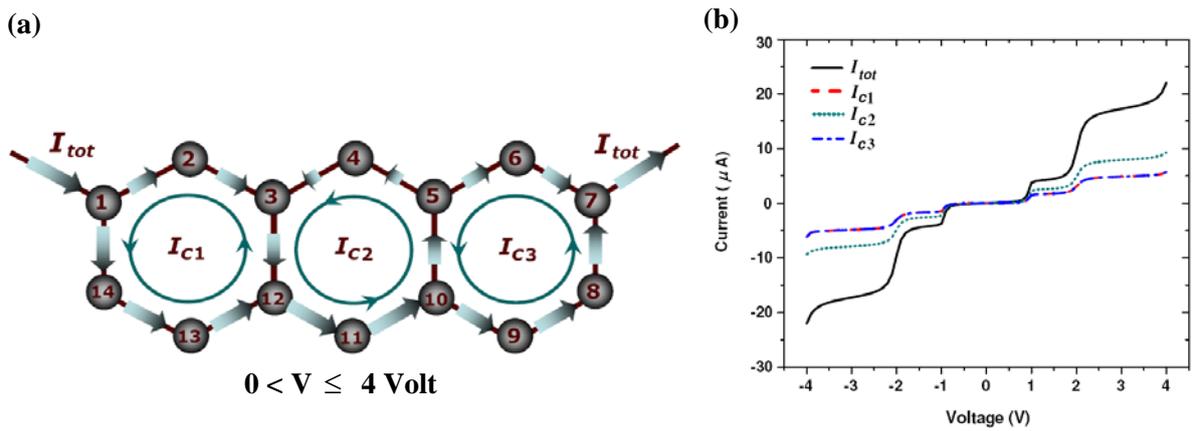

Fig. 12. (a) Current distribution pattern in anthracene. (b) Variation of net current I and circular currents ($I_{C1}, I_{C2}, I_{C3}$) with applied voltage, V.

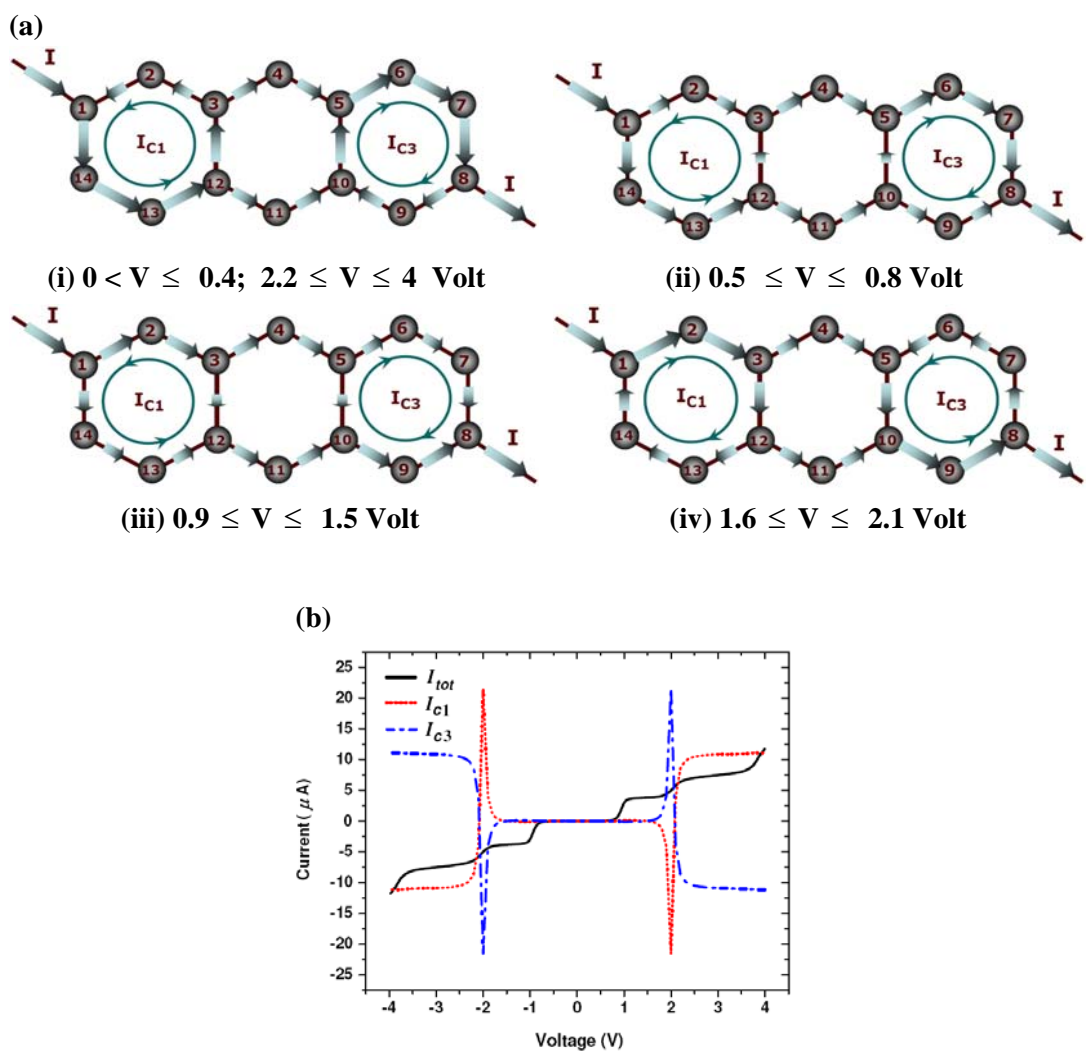

Fig. 13. (a) Internal current distribution in diagonally connected anthracene. (b) Current-voltage characteristics.

## 5. Concluding Remarks

We have investigated the phenomenon of circular currents in driven molecular wires characterized with loop structures, focusing on three issues. First, we have addressed the quantitative definition of a circular current and have suggested that a consistent and meaningful definition can be made by identifying this current as the source of the loop-induced magnetic field. Secondly, noticing that circular currents may be viewed as resulting from interference between carrier wavefunctions propagating along different pathways, we have studied their behavior under imposed decoherence and the implications of dephasing processes on the resulting magnetic fields. Finally, we have studied the circular current and the associated magnetic fields in simple tight binding models of several small molecular wire structures with loops – benzene, naphthalene, anthracene and azulene. Circular currents are found to be pervasive in driven molecular wires of this type, depending on junction geometry and voltage. As noted in previous studies we have found that for some structures and in certain ranges of imposed voltage circular currents can be much larger than the net current through the molecule, and the resulting magnetic fields can be considerable, e.g. ~ 0.23 T at 2V bias voltage in the model studied for meta-connected Benzene.

It is both interesting and important to consider the way such phenomena, so prominent in theoretical calculations, can be detected experimentally. Two routes to such observations may be considered. First is the spectral response of magnetic ions, placed on or near the ring, to the magnetic field which forms in their neighborhood, and secondly, the response of the magnetic moment developed on the molecule to an external magnetic field. These issues will be considered in a forthcoming paper.

**Appendix A**

In a current carrying steady state of a molecular ring driven by a voltage bias, different segments $\{j\}$ of the ring usually carry different currents $\{I_j\}$. Obviously, one can always redefine the segmental currents to be $\{I_j - I_c\}$ and assign to the ring a circular current $I_c$ that (a) does not affect the net current flowing into and out of the ring, and (b) adds to all bond currents. An additional criterion is needed to define $I_c$ uniquely. Here we suggest three alternative definitions based on the magnetic field induced by the current, and show that they all lead to the same assignment of $I_c$.

The circular current may be defined as the current component that induces (a) the magnetic flux threading the molecular ring, (b) the magnetic field at the center of the ring, and (c) the magnetic moment at the center of the ring.

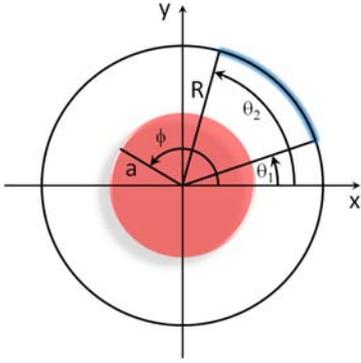

Fig. 14. Geometry used in the discussion of the magnetic properties of a ring current.

*Flux-based definition.* We start by calculating the magnetic flux threading an inner circle of radius $a$ (marked red in the online Fig. 14) due to a current carrying arc (thicker line, marked blue in the online figure). The flux is given by the following expression:

$$\phi_B = \int_S \vec{B} \cdot \hat{n}\, ds = \int_S (\vec{\nabla} \times \vec{A}) \cdot \hat{n}\, ds = \oint_c \vec{A} \cdot d\vec{l}_a . \qquad (A1)$$

Here, $\vec{B}$ is the induced magnetic field, $\hat{n} = (0,0,1)$ - is a unit vector normal to the surface of the ring, $ds$ - is a surface element, $\vec{A}$ - is the induced vector potential, and $d\vec{l}_a = (dx, dy, 0) = d(a\cos(\phi), a\sin(\phi), 0) = a(d\cos(\phi), d\sin(\phi), 0) = a(-\sin(\phi), \cos(\phi), 0)d\phi$ - is an infinitesimal line segment along the inner circle circumference. The surface integrals are taken over the full surface area of the inner circle and the line integral is taken along the circumference of this circle.

The vector potential at a point $a(\cos(\phi), \sin(\phi), 0)$ on the circumference of the inner circle induced by an infinitesimal segment of the current carrying arc at point $R(\cos(\theta), \sin(\theta), 0)$ can be calculated using the Biot-Savart Law:

$$d\vec{A} = \frac{\mu_0 I}{4\pi} \frac{d\vec{l}_R}{r} \tag{A2}$$

Here, $\mu_0 = 4\pi \times 10^{-7} NA^{-2} (Tm/A)$ is the magnetic constant, $I$ – is the current flowing through the arc, $d\vec{l}_R = R(-\sin(\theta), \cos(\theta), 0) d\theta$ - is an infinitesimal line segment along the current carrying arc, and

$$r = \sqrt{(R\cos(\theta) - a\cos(\phi))^2 + (R\sin(\theta) - a\sin(\phi))^2} = \sqrt{R^2 + a^2 - 2aR\cos(\phi - \theta)}$$

- is the distance between the current carrying segment and the point at which the vector potential is evaluated. Note that we use the standard convention by which a counter-clockwise current is taken to be positive.

Integrating over the full length of the arc we obtain:

$$\vec{A} = \frac{\mu_0 I R}{4\pi} \int_{\theta_1}^{\theta_2} \frac{(-\sin(\theta), \cos(\theta), 0)}{\sqrt{R^2 + a^2 - 2aR\cos(\phi - \theta)}} d\theta \tag{A3}$$

and the magnetic flux is now given by:

$$\phi_B = \oint_c \vec{A} \cdot d\vec{l}_a = \frac{\mu_0 I a R}{4\pi} \int_0^{2\pi} d\phi \int_{\theta_1}^{\theta_2} d\theta \frac{(-\sin(\theta), \cos(\theta), 0) \cdot (-\sin(\phi), \cos(\phi), 0)}{\sqrt{R^2 + a^2 - 2aR\cos(\phi - \theta)}} =$$

$$\frac{\mu_0 I a R}{4\pi} \int_0^{2\pi} d\phi \int_{\theta_1}^{\theta_2} d\theta \frac{\sin(\theta)\sin(\phi) + \cos(\theta)\cos(\phi)}{\sqrt{R^2 + a^2 - 2aR\cos(\phi - \theta)}} = \frac{\mu_0 I a R}{4\pi} \int_{\theta_1}^{\theta_2} d\theta \int_0^{2\pi} d\phi \frac{\cos(\phi - \theta)}{\sqrt{R^2 + a^2 - 2aR\cos(\phi - \theta)}}$$

(A4)

Since the integrand is a periodic function of the angles difference, the double integral can be replaced by a single integral. To show this we change variables to $\eta \equiv \phi - \theta$ to obtain:

$$\int_0^{2\pi} d\phi \frac{\cos(\phi - \theta)}{\sqrt{R^2 + a^2 - 2aR\cos(\phi - \theta)}} = \int_{-\theta}^{2\pi - \theta} d\eta \frac{\cos(\eta)}{\sqrt{R^2 + a^2 - 2aR\cos(\eta)}}, \tag{A5}$$

This leads to

$$\varphi_B = \frac{\mu_0 I a R}{4\pi} \int_{\theta_1}^{\theta_2} d\theta \int_{-\theta}^{2\pi - \theta} d\eta \frac{\cos(\eta)}{\sqrt{R^2 + a^2 - 2aR\cos(\eta)}} = \frac{\mu_0 I a R (\theta_2 - \theta_1)}{4\pi} \int_0^{2\pi} d\eta \frac{\cos(\eta)}{\sqrt{R^2 + a^2 - 2aR\cos(\eta)}}$$

$$= \frac{\mu_0 I k l_j}{4\pi} \int_0^{2\pi} d\eta \frac{\cos(\eta)}{\sqrt{1 + k^2 - 2k\cos(\eta)}}$$

(A6)

Here the second equality results from the fact that the integration is taken over a full period of the periodic integrand, $l_j$ is the length of the current carrying arc, and $k \equiv \dfrac{a}{R}$. The remaining integral can be expressed in terms of elliptic integrals in the following form:

$$\int_0^{2\pi} d\eta \frac{\cos(\eta)}{\sqrt{1+k^2-2k\cos(\eta)}} = \left[\frac{(k-1)}{k} E\left(\frac{\eta}{2}, -\frac{4k}{(k-1)^2}\right) + \frac{k^2+1}{k(1-k)} F\left(\frac{\eta}{2}, -\frac{4k}{(k-1)^2}\right)\right]_0^{2\pi} =$$

$$\frac{(k-1)}{k} E\left(\pi, -\frac{4k}{(k-1)^2}\right) + \frac{k^2+1}{k(1-k)} F\left(\pi, -\frac{4k}{(k-1)^2}\right) = \frac{2(k-1)}{k} E\left(-\frac{4k}{(k-1)^2}\right) + 2\frac{k^2+1}{k(1-k)} K\left(-\frac{4k}{(k-1)^2}\right)$$

(A7)

Here, $F(\phi, k) = \int_0^{\phi} \dfrac{d\theta}{\sqrt{1-k^2\sin^2(\theta)}}$ is the incomplete elliptic integral of the first kind,

$E(\phi, k) = \int_0^{\phi} \sqrt{1-k^2\sin^2(\theta)}\, d\theta$ is the incomplete elliptic integral of the second kind,

$K(k) = \int_0^{\pi/2} \dfrac{d\theta}{\sqrt{1-k^2\sin^2(\theta)}}$ is the complete elliptic integral of the first kind, and

$E(k) = \int_0^{\pi/2} \sqrt{1-k^2\sin^2(\theta)}\, d\theta$ is the complete elliptic integral of the second kind. Eqs. (A6)-(A7) then lead to the following result for the magnetic flux induced by the current carrying arc:

$$\Phi_B = \frac{\mu_0 I k l_j}{4\pi} \int_0^{2\pi} d\eta \frac{\cos(\eta)}{\sqrt{1+k^2-2k\cos(\eta)}} = \frac{\mu_0 I l}{2\pi}\left[(k-1)E\left(-\frac{4k}{(k-1)^2}\right) - \frac{k^2+1}{(k-1)} K\left(-\frac{4k}{(k-1)^2}\right)\right]$$

(A8)

The magnetic flux induced by both arms in the inner circle of radius $a$ (Fig. 13) is thus given by:

$$\Phi_B = \frac{\mu_0}{2\pi}\left[(k-1)E\left(-\frac{4k}{(1-k)^2}\right) - \left(\frac{1+k^2}{k-1}\right) K\left(-\frac{4k}{(1-k)^2}\right)\right](I_1 l_1 + I_2 l_2) \quad (A9)$$

Note that this expression diverges when $k=1$, that is, $a=R$, as is well known for a loop current of zero width. However the form (A9) is sufficient to define the transverse and circular current components associated with the current distribution in Fig. 1. Defining $I_j^{tr} = I_j - I_c$, $j = 1, 2$, we require that the transverse current components $I_j^{tr}$ satisfy that the magnetic flux vanishes for any choice of inner radius $a$. Using (A9) this translates to

$$I_1^{tr} l_1 + I_2^{tr} l_2 = 0. \quad (A10)$$

In addition, since the circular current does not contribute to the total current, the sum of the transverse current components on both arms should produce the total current, i.e.,

$$I_2^{tr} - I_1^{tr} = I_{tot}. \tag{A11}$$

Eqs. (A10) and (A11) now lead to

$$I_1^{tr} = -I_{tot}\frac{l_2}{L} \quad ; \quad I_2^{tr} = I_{tot}\frac{l_1}{L} \tag{A12}$$

where $L = l_1 + l_2 = 2\pi R$, and

$$I_c = I_1 - I_1^{tr} = I_2 - I_2^{tr} = \frac{I_1 l_1 + I_2 l_2}{L} \tag{A13}$$

*Magnetic field and magnetic moment based definitions.* Obviously, any quantity whose dependence on the current distribution on the loop enters through proportionality to $I_1 l_1 + I_2 l_2$ will vanish together with the magnetic flux. Consider for example the magnetic field at the center of the ring. The magnetic field produced by a current carrying arc at the center of the ring can be calculated from the Biot-Savart expression

$$d\vec{B}^c = \frac{\mu_0 I}{4\pi}\frac{d\vec{l}_R \times \vec{r}}{r^3}, \tag{A14}$$

Taking, as before, the ring to be in the xy plane with its center at the origin, we have $\vec{r} = (0-x, 0-y, 0) = (-R\cos(\theta), -R\sin(\theta), 0)$ and $r \equiv |\vec{r}| = R$. Thus,

$$d\vec{l}_R \times \vec{r} = \begin{vmatrix} \hat{x} & \hat{y} & \hat{z} \\ -R\sin(\theta)d\theta & R\cos(\theta)d\theta & 0 \\ -R\cos(\theta) & -R\sin(\theta) & 0 \end{vmatrix} = (0, 0, R^2\sin^2(\theta)d\theta + R^2\cos^2(\theta)d\theta) = (0, 0, R^2 d\theta) \quad \backslash$$

(A15)

And from (A12), the corresponding contribution to the magnetic field at the ring center is

$$d\vec{B}^c = \frac{\mu_0 I}{4\pi}\frac{(0,0,1)R^2 d\theta}{R^3} = \frac{\mu_0 I}{4\pi R}(0,0,1)d\theta. \tag{A16}$$

Integrating over the angle $\theta$ that defined the arc gives the arc contribution in the form

$$\vec{B}^c = \frac{\mu_0 I}{4\pi R}(0,0,1)(\theta_2 - \theta_1) = \frac{\mu_0 I l_j}{4\pi R^2}(0,0,1) \tag{A17}$$

with $l_j = R(\theta_1 - \theta_2)$ being the length of the arc. Summing over all arcs with their corresponding currents (Fig. 1) yields the field at the ring center

$$\vec{B}^c = \frac{\mu_0 (I_1 l_1 + I_2 l_2)}{2\pi L}(0,0,1) \tag{A18}$$

Defining the transverse current as that component of the current that nulls this field obviously leads the same result as before.

Next consider the magnetic moment at the center of the ring. The contribution to this moment from a given arc element is

$$d\vec{m}^c = \frac{I}{2}\left(\vec{r} \times d\vec{l}_R\right) \tag{A19}$$

Using as before, $\vec{r} = -R(\cos(\theta), \sin(\theta), 0)$ and $d\vec{l}_R = R(-\sin(\theta), \cos(\theta), 0)d\theta$ leads to

$$d\vec{m}^c = -\frac{IR^2}{2}(0,0,1)d\theta \tag{A20}$$

Integrating over the arc yields

$$\vec{m}^c = -\frac{IR^2}{2}(\theta_2 - \theta_1)(0,0,1) = -\frac{1}{2}RIl_j(0,0,1). \tag{A21}$$

Summing over the two arcs in Fig. 1 then yields

$$\vec{m}^c = -(R/2)(I_1 l_1 + I_2 l_2)(0,0,1) \tag{A22}$$

with the same implications as before on the definition of the transverse and circular current components.

The above considerations can be generalized further in two important ways. First, if the circular rings includes several segments of lengths and currents $l_j$ and $I_j$, respectively, the magnetic flux expression, Eq. (A9) becomes

$$\Phi_B = \frac{\mu_0}{2\pi}\left[(k-1)E\left(-\frac{4k}{(1-k)^2}\right) - \left(\frac{1+k^2}{k-1}\right)K\left(-\frac{4k}{(1-k)^2}\right)\right]\sum_j I_j l_j \tag{A23}$$

Similarly, the magnetic field at the ring center and the magnetic moment also become proportional to $\sum_j I_j l_j$. The transverse currents, $I_j - I_c$, should null these magnetic effects, i.e.

$$\sum_j (I_j - I_c)l_j = 0 \tag{A24}$$

Implying that

$$I_c = \frac{\sum_j I_j l_j}{L} \tag{A25}$$

and

$$I_j^{tr} = I_j - I_c = \frac{\sum_{j'}(I_j - I_{j'})l_{j'}}{L} \tag{A26}$$

These results have made it possible for us the uniquely defined the circular and transverse currents on different rings of polycyclic molecules (Section 4).

Second, if instead of a perfect circle we have a polygon of $N$ equal sides of length $b$, the contribution of segment of $n_j$ sides carrying a current $I_j$ ($\sum_j n_j = N$) to the magnetic property under consideration is proportional to $bn_j I_j$, so the total magnetic property is proportional to $\sum_j n_j I_j$. This leads to

$$I_c = \frac{\sum_j I_j n_j}{N} \quad \text{and} \quad I_j^{tr} = \frac{\sum_{j'}(I_j - I_{j'})n_{j'}}{N} \tag{A27}$$

**Acknowledgements.** We thank Prof. Haim Diamant and Prof. Shahar Hod for helpful discussions. This research of A.N. is supported by the Israel Science Foundation, the Israel-US Binational Science Fundation, the European Science Council (FP7 /ERC grant no. 226628) and the Israel – Niedersachsen Research Fund. O.H. acknowledges the support of the Israel Science Foundation under grant # 1313/08. D.R. thanks the Center for Nanoscience and Nanotechnology at Tel-Aviv University for supporting post doctoral fellowship.